\documentclass[preprint,showpacs,titlepage,aps,prd,tightenlines,
amsmath,byrevtex,nofootinbib]{revtex4}

\usepackage{graphicx}

\begin{document}

\preprint{hep-ph/0404142}

\title{Unity of CP and T Violation in Neutrino Oscillations}

\author{Martin Blom\protect\footnote{
Also at: Division of Mathematical Physics, Department of Physics,
Royal Institute of Technology (KTH), AlbaNova University Center,
Roslagstullsbacken 11, 106 91 Stockholm, Sweden}
}
\email{martin@phys.metro-u.ac.jp}
\author{Hisakazu Minakata}
\email{minakata@phys.metro-u.ac.jp}
\affiliation{Department of Physics, Tokyo Metropolitan University \\
1-1 Minami-Osawa, Hachioji, Tokyo 192-0397, Japan}

\date{\today}

\vglue 1.4cm
\begin{abstract}
In a previous work a simultaneous $P$- $CP[P]$ and $P$- $T[P]$ 
bi-probability plot was proposed as a useful tool for 
unified graphical description of CP and T violation in neutrino 
oscillation. The diamond shaped structure of the plot 
is understood as a consequence of the approximate CP-CP and 
the T-CP relations obeyed by the oscillation probabilities. 
In this paper, we make a step forward toward deeper understanding 
of the unified graphical representation by showing that these 
two relations are identical in its content, suggesting 
a truly unifying view of CP and T violation in neutrino 
oscillations. We suspect that the unity reflects the 
underlying CPT theorem.
We also present calculation of corrections to the CP-CP 
and the T-CP relations to leading order in 
$\Delta m^2_{21} / \Delta m^2_{31}$ and $s^2_{13}$.

\end{abstract}


\maketitle


\section{Introduction}

Exploring leptonic CP and T violation is one of the most 
challenging endeavors in particle physics. 
Confirming (or refuting) unsuppressed CP violation 
analogous to that in the quark sector must shed light on 
deeper understanding of lepton-quark correspondence, 
the concept whose importance was recognized 
early in sixties \cite{nagoya}. 
We should note, however, that it is only after the KamLAND 
experiment \cite{KamLAND} which confirmed the MSW 
large mixing angle (LMA) solution \cite{wolfenstein,MSW} 
of the solar neutrino problem that we can talk about 
detecting CP or T violation in an experimantally  
realistic setting. 
An almost maximal mixing of $\theta_{23}$ discovered by 
the atmospheric neutrino observation by Super-Kamiokande 
\cite{SKatm}, which broke new ground in the field of 
research, also greatly encourages attempts toward 
measuring the leptonic Kobayashi-Maskawa phase $\delta$.

Yet, we might have the last impasse to observing leptonic 
CP violation, a too small value of $\theta_{13}$, which lives 
in the unique unexplored (1-3) sector of the MNS matrix \cite{MNS}.
Currently, it is bounded from above by a modest constraint 
$\sin^2{2\theta_{13}} \leq 0.15-0.25$ obtained by the Chooz 
reactor experiment \cite{CHOOZ}. 
Toward removing the last impasse, two different methods for 
measuring $\theta_{13}$ are proposed and materialized into a 
number of concrete experimental programs. 
The first is the measurement of appearance probability 
$P(\nu_{\mu} \rightarrow \nu_{e})$ in long-baseline (LBL) experiments 
using accelerator neutrino beam, being and to be performed by the 
ongoing \cite{MINOS,OPERA} and the next generation projects 
\cite{JPARC,NuMI,SPL}. 
The second is the reactor measurement of $\theta_{13}$. It is a 
pure measurement of $\theta_{13}$ independent of other oscillation 
parameters, $\delta$ and $\theta_{23}$, and thus will play a r{\^o}le
complementary to the LBL experiments \cite{MSYIS}.
This property is expected to help resolving the parameter degeneracy 
\cite{Burguet-C,MNjhep01,KMN02,octant,BMW1,MNP2} related to 
$\theta_{23}$ \cite{MSYIS}.
A spur of experimental projects that occurred over the globe for  
the relatively new opportunity are now summarized in the 
White Paper Report \cite{reactor_white}.

If such challenges are blessed by nature we will be able to 
proceed to measuring the leptonic CP or T violating phase $\delta$. 
The relatively large value of $\theta_{13}$ will allow us to measure it 
via $\nu_{e}$ and $\bar{\nu}_{e}$ appearance measurement using 
conventional superbeam experiments, whose idea may be traced back to 
\cite{superbeam}. 
Feasible experimental programs for such appearance measurement 
with upgraded beams as well as detectors are proposed. 
See e.g.,  \cite{JPARC,Hyper-K} for the JPARC-Hyper-Kamiokande project 
and \cite{NuMI} for NO$\nu$A. 
It is also proposed that a fast search for CP violation can be performed 
by combining neutrino mode operation of such experiments with high 
statistics reactor measurement of $\theta_{13}$ \cite{reactorCP}.

If $\theta_{13}$ is too small to be seen in the above experiments an 
entirely new strategy is called for. We will probably need more aggressive 
approach with ambitious beam technologies, 
neutrino factory \cite{nufact} or beta beam \cite{beta} or even both.
Here also, vigorous world-wide activities for developing beam and target 
technologies as well as studying physics capabilities are underway 
\cite{nufact_eu, nufact_us,beta2}. 
Intense neutrino beam from a muon storage ring and the clean 
background for wrong sign muon detection are expected to lead to 
an enormous sensitivity of $\theta_{13}$ up to $\sim$ 1 degeree. 
Enriched by golden ($\nu_{e} \rightarrow \nu_{\mu}$) \cite{golden} as well as 
silver ($\nu_{e} \rightarrow \nu_{\tau}$) \cite{silver} channels, it will be able to 
resolve all the parameter degeneracies, as claimed in \cite{donini}. 
See \cite{NOVE_mina} for a review of old and new ideas on 
how to measure leptonic CP violation.

How does T violation measurement fit into the scene? 
To our understanding it will probably come later than CP violation 
measurement because the measurement is more difficult to carry out. 
In neutrino factory it requires electron charge identification which 
is highly nontrivial, if not impossible. The beta beam, if build, 
would give us an ideal apparatus because it can deliver a pure 
$\nu_e$ beam which comes from decaying radioactive nuclei. 
By combining with superbeam (or neutrino factory) measurement of 
$P(\nu_{\mu} \rightarrow \nu_{e})$ it will provides us 
a unique opportunity for exploring leptonic T violation.

Keeping in mind the scope of experimental realization of CP and T violation 
measurement in the future, 
we discuss in this article a unified view of leptonic CP and T violation, 
one of the most fundamental problems in particle physics. 
We hope that our discussion is illuminating and contributes to deeper 
understanding of the problem. 
In this paper we use, except for in Appendix, 
the standard notation of the MNS matrix \cite{PDG}.

\section{CP and T violation in neutrino oscillation}

It has been known for a long time that CP and T conservation 
are intimately related to each other by the CPT theorem. 
For neutrino oscillation in vacuum the invariance leads 
to a relation between neutrino and 
antineutrino  oscillation probabilities 
\begin{eqnarray}
P(\nu_{\alpha} \rightarrow \nu_{\beta}; \delta) =
P(\bar{\nu}_{\beta} \rightarrow \bar{\nu}_{\alpha}; \delta). 
\end{eqnarray}
Then, the question might be ``if there exists analogous 
relation in neutrino oscillation in matter?''.
It was shown in \cite{MNP1} that indeed there exists such a 
relationship, 
\begin{eqnarray}
P(\nu_{\alpha} \rightarrow \nu_{\beta};
\Delta m^2_{31},\Delta m^2_{21}, \delta, a)
&=&
P(\bar{\nu}_{\beta} \rightarrow \bar{\nu}_{\alpha};
\Delta m^2_{31},\Delta m^2_{21}, \delta, -a), 
\label{CPT-ids}
\end{eqnarray}
which comes from the classical time reversal and
the complex conjugate of the neutrino evolution equation assuming 
that the matter profile is symmetric about the mid-point
between production and detection.
Let us call (\ref{CPT-ids}) the CPT relation in matter.
Here, $a = 2 \sqrt{2} G_F N_e E$ is the fundamental quantity 
which is related to neutrino's index of refraction 
in matter \cite{wolfenstein} 
with $G_F$ being the Fermi constant, $E$ neutrino energy, 
and $N_e(x)$ an electron number density in the earth. 
The mass squared difference of neutrinos 
is defined as $\Delta m^2_{ij} \equiv m^2_i - m^2_j$ where
$m_i$ is the mass of the $i$th eigenstate.

There is an immediate consequence of the CPT relation in matter, 
Eq.~(\ref{CPT-ids}). 
If we define $\Delta P_{CPT}$ as 
\begin{eqnarray}
\Delta P_{CPT} \equiv
P(\nu_{\alpha} \rightarrow \nu_{\beta};
\Delta m^2_{31},\Delta m^2_{21}, \delta, a) - 
P(\bar{\nu}_{\beta} \rightarrow \bar{\nu}_{\alpha};
\Delta m^2_{31},\Delta m^2_{21}, \delta, a), 
\label{DeltaPCPT} 
\end{eqnarray}
then, $\Delta P_{CPT}$ is an odd function of $a$.
The property may be used to formulate the method for detecting 
extrinsic CPT violation in neutrino oscillation due to 
matter effect \cite{ohlsson}.

Do the CPT relation and various other relationships between 
oscillation probabilities give a unified picture of CP 
and T violation in neutrino oscillation in matter? 
In this article we argue that the answer is indeed yes.
Although our argument in this paper is based on the line 
of thought in \cite{MNP1}, we believe that we made a step 
forward from the previous work.

\section{Unified graphical representation of CP and T Violation}

Toward the goal of this paper, let us introduce a graphical 
representation of the characteristic features of neutrino 
oscillations relevant for leptonic CP violation \cite{MNjhep01}. 
For simplicity of notations let us define the symbols for 
CP and T conjugate probabilities, 
$CP[P] \equiv P(\bar{\nu}_{\alpha} \rightarrow \bar{\nu}_{\beta})$ and 
$T[P] \equiv P(\nu_{\beta} \rightarrow \nu_{\alpha})$, 
for a given probability 
$P(\nu_{\alpha} \rightarrow \nu_{\beta})$. 
It is the CP trajectory diagram in the $P$-$CP[P]$ bi-probability 
space, which can be extended to incorporate the $P$-$T[P]$ 
bi-probability plot \cite{MNP1}.

\begin{figure}[htbp]
\begin{center}
\includegraphics[width=0.7\textwidth]{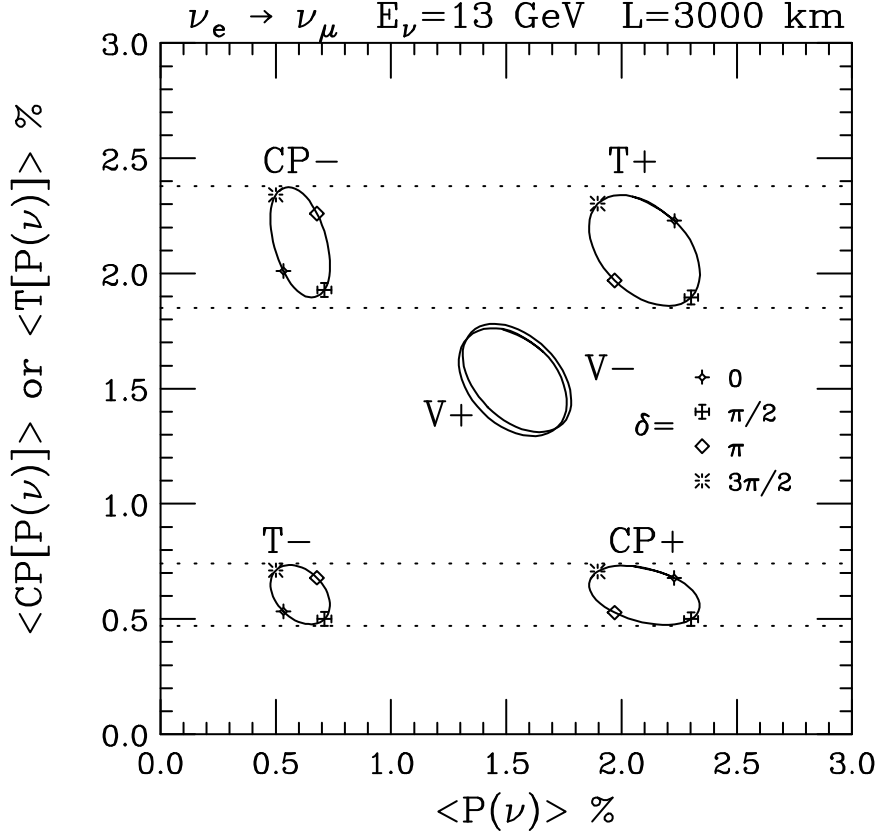}
\end{center}
\caption{
A simultaneous $P$- $T[P]$ and $P$- $CP[P]$ bi-probability plot 
with experimental parameters
corresponding to the baseline distance and about twice the 
optimal energy corresponding to maximal 
exhancement of T violating effect \protect\cite{PW01}.
Notice the difference between movement of the direction in 
$P$- $T[P]$ and $P$- $CP[P]$ plot; they are orthogonal to 
each other for reasons explained in the text.
The figure is the same as Fig.1 of \cite{MNP1} apart from that 
we have changed the convention of $\delta$ to the standard one 
used in the text of the paper. The convention is employed by almost 
everybody who works in the field (see e. g., \cite{Burguet-C,BMW1}) 
including all our previous works, \cite{MNP1,MNjhep01,MNP2,MNP3}, 
but is different from that of \cite{PDG}. Although the convention of the 
MNS matrix is the same, the latter takes a convention such that $U$ in 
the neutrino evolution equation (\ref{evolution1}) is replaced by 
$U^{*}$.
}
\label{PTP}
\end{figure}

Given two observables $P$ and $CP[P]$, you can draw a dot 
in $P$- $CP[P]$ space, and it becomes a closed ellipse when 
$\delta$ is varied.
In Fig.~\ref{PTP} the ellipses labeled $V_{\pm}$ are the ones 
for vacuum oscillation probabilities where the subscripts $\pm$
denote the sign of $\Delta m^2_{31}$. When the matter effect 
is turned on they split into two ellipses labeled 
$CP_{\pm}$ in the CP bi-probability plot and into 
$T_{\pm}$ in the T bi-probability plot, both of which are 
simultaneously depicted in Fig.~\ref{PTP}.

Let us first focus on the $P$- $CP[P]$ bi-probability plot. 
We first note that the oscillation probability 
$ P(\nu_{\alpha} \rightarrow \nu_{\beta})$
can be written on very general ground 
(for $\alpha = e$, $\beta = \mu, \tau$, or vice versa)\footnote{
In the $\nu_{\mu} - \nu_{\tau}$ channel, there exist 
$\sin{2\delta}$ and $\cos{2\delta}$ terms in the oscillation probability 
and the bi-probability diagrams are no more elliptic.
}
as \cite{KTY,KTY2}
\begin{eqnarray}
P(\nu_{\alpha} \rightarrow \nu_{\beta}) =
A \cos{\delta} + B \sin{\delta} + C
\label{kty}
\end{eqnarray}
where $A$, $B$, and $C$ are functions of $\Delta m^2_{31}$, 
$\Delta m^2_{21}$ and $a$. 
(Of course, the previously obtained approximate formulas 
do have such form. See, e.g., \cite{AKS,MNprd97,golden}.)
It is nothing but (\ref{kty}) 
that guarantees the elliptic nature of the trajectories.
Then, one can show that the lengths of major and minor axes 
(the ``polar'' and ``radial'' thickness of the ellipses) 
represent the size of the $\sin{\delta}$ and the $\cos{\delta}$ 
terms, respectively,  
whereas the distance between two ellipses with positive 
and negative $\Delta m^2_{31}$ displays the size of the 
matter effect \cite{MNjhep01}. 
Finally, the distance to the center of the ellipse from 
the origin is essentially given by $\sin^2{2 \theta_{13}}$. 
Notice that all the features of the bi-probability plot 
except for distance between $\Delta m^2_{31}=\pm$ ellipses 
are essentially determined by the vacuum parameters in 
setting of $E$ and $L$ relevant for the superbeam experiments. 
Therefore, one can easily guess how it looks 
like in the other experimental settings. 
As indicated in Fig.~\ref{PTP} 
the CP violating and CP conserving effects of 
$\delta$ are comparable in size with the matter effect even 
at such high energy and long baseline.

We notice that in the $P$- $T[P]$ bi-probability plot the 
matter effect splits the vacuum ellipses $V_{\pm}$ in quite a 
different way from the $P$- $CP[P]$ bi-probability plot. 
It is because the T violating measure $\Delta P_{T}$, 
which is given by 
\begin{eqnarray}
\Delta P_{T} &\equiv&  
P(\nu_{\alpha} \rightarrow \nu_{\beta}; \Delta m^2_{13}, \delta, a) - 
P(\nu_{\beta} \rightarrow \nu_{\alpha}; \Delta m^2_{13}, \delta, a) 
\nonumber \\
&=&
2B \sin{\delta}
\label{DeltaT}
\end{eqnarray} 
for symmetric matter profiles, vanishes at $\delta = 0$.
Therefore, the T (or CP) conserving point must remain 
on the diagonal line in $P$- $T[P]$ bi-probability plot.

Equation~(\ref{DeltaT}) stems from the fact that 
the coefficients except for $B$ are symmetric under the interchange 
$\alpha \leftrightarrow \beta$. 
Therefore, if $\Delta P_{T} \neq 0$, then $\delta \neq 0$ 
even in matter. The matter effect cannot create a fake 
T violation for symmetric matter profiles \cite{petcov}. 
For modifications which occur for asymmetric matter profiles, see e.g., 
\cite{miura,gouvea,munichT}.

Notice that the matter effect cannot create fake T violation, it does modify  
the coefficient $B$ in Eq.~(\ref{DeltaT}), whose feature is 
made transparent in \cite{PW01}. Among other things, it was shown 
in \cite{PW01} that the matter effect can enhance the 
T asymmetry up to a factor of 1.5.  Other earlier references on T violation 
in neutrino oscillation include \cite {kuo-panta,toshev,yasuda,koike-sato}.

\section{Diamond shaped structure of CP and T bi-probability plot}

In view of Fig.~(\ref{PTP}) we notice a remarkable feature of 
the simultaneous $P$-$CP[P]$ and $P$-$T[P]$ diagrams; it  
has a square or diamond shape.
%
The diamond shaped structure of combined $P$-$CP[P]$
and $P$-$T[P]$ diagrams can be understood by the 
following two relations which are called the CP-CP and the 
T-CP relations in \cite{MNP1}. Their precise statements are:

\vskip 0.2cm

\noindent
CP-CP relation:
\begin{eqnarray}
P(\nu_{e} \rightarrow \nu_{\mu};
\Delta m^2_{31},\Delta m^2_{21}, \delta,a)
&=&
P(\bar{\nu}_{e} \rightarrow \bar{\nu}_{\mu};
-\Delta m^2_{31},-\Delta m^2_{21},\delta, a) \nonumber \\
& \approx &
P(\bar{\nu}_{e} \rightarrow \bar{\nu}_{\mu};
-\Delta m^2_{31},+\Delta m^2_{21},\pi + \delta, a). 
\label{CP-CP}
\end{eqnarray}

\vskip 0.2cm

\noindent
T-CP relation:
\begin{eqnarray}
P(\nu_{\mu} \rightarrow \nu_{e};
\Delta m^2_{31},\Delta m^2_{21}, \delta, a)
&=&
P(\bar{\nu}_{e} \rightarrow \bar{\nu}_{\mu};
-\Delta m^2_{31},-\Delta m^2_{21},2 \pi - \delta, a) \nonumber  \\
& \approx &
P(\bar{\nu}_{e} \rightarrow \bar{\nu}_{\mu};
-\Delta m^2_{31},+\Delta m^2_{21}, \pi - \delta, a ).
\label{T-CP}
\end{eqnarray}
These relations are meant to be valid in leading order in 
$\Delta m^2_{21}/\Delta m^2_{31}$, i.e., in zeroth order in 
$\delta$-independent and to first order in $\delta$-dependent 
terms, respectively.

Roughly speaking, the 
CP-CP relation guarantees that the locations of the first and 
the third bases are approximately symmetric under reflection with 
respect to the diagonal line in $P$-$CP[P]$ space, whereas 
the T-CP relation guarantees that the ordinates of the 
$T_{\pm}$ ellipse are approximately the same as those of 
$CP_{\mp}$. 
Of course, one has to specify the values of the CP phase $\delta$ to 
make the relationship precise, and that is why the change in $\delta$ 
is involved between the RHS and the LHS of Eqs.(\ref{CP-CP}) and (\ref{T-CP}).

A rough sketch of the argument given in \cite{MNP1} is as follows.
The first equality in (\ref{CP-CP}) is obvious by noticing 
\begin{eqnarray}
P(\nu_{e} \rightarrow \nu_{\mu};
\Delta m^2_{31}, \Delta m^2_{21}, \delta, a) = 
P(\nu_{e} \rightarrow \nu_{\mu};
- \Delta m^2_{31}, - \Delta m^2_{21}, - \delta, -a) 
\label{ccid}
\end{eqnarray}
which follows from 
the fact that a complex conjugate of the neutrino evolution equation 
gives the same oscillation probability, and the simple relation 
(see e.g., \cite{MNprd97}) 
\begin{eqnarray}
P(\bar{\nu}_{e} \rightarrow \bar{\nu}_{\mu};
\Delta m^2_{31}, \Delta m^2_{21}, \delta, a) =
P(\nu_{e} \rightarrow \nu_{\mu};
\Delta m^2_{31}, \Delta m^2_{21}, - \delta, -a)
\end{eqnarray}
Then, the second approximate equality follows to leading order in 
$\Delta m^2_{21} / \Delta m^2_{31}$ 
after appropriate shift of $\delta$ which takes care of the 
sign change in $\delta$ dependent terms.

For the T-CP relation the first equality in (\ref{T-CP}) can be derived 
by using (\ref{ccid}) in the CPT relation in matter (\ref{CPT-ids}). 
Then, the second approximate equality holds for small 
$\Delta m^2_{21} / \Delta m^2_{31}$ with the same 
adjustment of the phase $\delta$.

\section{Equivalence between the CP-CP and the T-CP relations}

We now point out that the the CP-CP and the T-CP relations are 
equivalent to each other in their physics contents. 
Roughly speaking, the T-CP relation is ``T conjugate'' of 
the CP-CP relation. 
It reflects the relationships among various neutrino 
oscillation probabilities discussed in the previous section.
Their equivalence again testifies for the unity of CP and T 
violation in neutrino oscillations.

We present the proof of the above statement through the computation 
of the corrections to the CP-CP and the T-CP relations. 
Let us define for clarity of notations the deviation from 
the CP-CP and the T-CP relations as
\begin{eqnarray}
\Delta P_{CP-CP} &\equiv&
P(\nu_{e} \rightarrow \nu_{\mu};
\Delta m^2_{31},\Delta m^2_{21}, \delta, a) - 
P(\bar{\nu}_{e} \rightarrow \bar{\nu}_{\mu};
-\Delta m^2_{31},+\Delta m^2_{21},\pi + \delta, a), 
\label{DeltaPCP} \\
\Delta P_{T-CP} &\equiv&
P(\nu_{\mu} \rightarrow \nu_{e};
\Delta m^2_{31},\Delta m^2_{21}, \delta, a) -
P(\bar{\nu}_{e} \rightarrow \bar{\nu}_{\mu};
-\Delta m^2_{31},+\Delta m^2_{21}, \pi - \delta, a).
\label{DeltaPT}
\end{eqnarray}
Because of the first equality in (\ref{CP-CP}), $\Delta P_{CP-CP}$ 
can be written as 
\begin{eqnarray}
\Delta P_{CP-CP} = 
P(\nu_{e} \rightarrow \nu_{\mu};
\Delta m^2_{31},\Delta m^2_{21}, \delta, a) - 
P(\nu_{e} \rightarrow \nu_{\mu};
+\Delta m^2_{31}, -\Delta m^2_{21}, \pi + \delta, a) 
\label{DeltaPCP2}
\end{eqnarray}
On the other hand, by using the first equality in (\ref{CP-CP}) 
and the relation 
\begin{eqnarray}
P(\nu_{\mu} \rightarrow \nu_{e};
\Delta m^2_{31},\Delta m^2_{21}, \delta, a) = 
P(\nu_{e} \rightarrow \nu_{\mu};
\Delta m^2_{31}, \Delta m^2_{21}, -\delta, a), 
\label{Treversal}
\end{eqnarray}
which is valid for symmetric matter density profiles, 
$\Delta P_{T-CP}$ can be cast into the form 
\begin{eqnarray}
\Delta P_{T-CP} =
P(\nu_{e} \rightarrow \nu_{\mu};
\Delta m^2_{31}, \Delta m^2_{21}, -\delta, a) - 
P(\nu_{e} \rightarrow \nu_{\mu};
+\Delta m^2_{31}, -\Delta m^2_{21}, \pi - \delta, a). 
\label{DeltaPT2}
\end{eqnarray}
Therefore, it holds that 
\begin{eqnarray}
\Delta P_{T-CP}(\delta) =
\Delta P_{CP-CP}(-\delta) 
\hspace*{7mm} 
(\mbox{mod.} 2\pi)
\label{DeltaCPT}
\end{eqnarray}
Namely, the relation (\ref{DeltaPT}) is the 
T-conjugate of  (\ref{DeltaPCP}). 
This completes the proof that 
the CP-CP and the T-CP relations are identical in their content.

In passing, we note the following: 
It was noted in \cite{MNjhep01} that there exists an approximate 
symmetry in the vacuum oscillation probability under the 
simultaneous transformations  
$\Delta m^2_{31} \rightarrow -\Delta m^2_{31}$ and 
$\delta \rightarrow \pi-\delta$ which explains almost overlap of 
$V_{+}$ and $V_{-}$ trajectories as in Fig.~\ref{PTP}. 
A generalization of the approximate symmetry into the case with 
matter effect has been obtained \cite{MNP1}, 
\begin{equation}
P(\nu_{\alpha} \rightarrow \nu_{\beta};
~\Delta m^2_{31},~\Delta m^2_{21}, \delta,a)
\approx
P(\nu_{\alpha} \rightarrow \nu_{\beta};-\Delta m^2_{31},\Delta m^2_{21},
\pi - \delta, -a), 
\label{flipsym2}
\end{equation}
from which the CP-CP and the T-CP relations also follow.
Clearly, the correction to the approximate symmetry is also 
related to $\Delta P_{CP-CP}$. To show this, we define the 
difference $\Delta P_{flip}$ between the LHS and the RHS of 
(\ref{flipsym2}), 
\begin{equation}
\Delta P_{flip} \equiv
P(\nu_{\alpha} \rightarrow \nu_{\beta};
~\Delta m^2_{31},~\Delta m^2_{21}, \delta,a) - 
P(\nu_{\alpha} \rightarrow \nu_{\beta};-\Delta m^2_{31},\Delta m^2_{21},
\pi - \delta, -a).  
\label{DeltaPflip}
\end{equation}
Using the frequently used identity, one can show that 
\begin{equation}
\Delta P_{flip} = 
P(\nu_{\alpha} \rightarrow \nu_{\beta};
~\Delta m^2_{31},~\Delta m^2_{21}, \delta,a) - 
P(\nu_{\alpha} \rightarrow \nu_{\beta}; +\Delta m^2_{31}, -\Delta m^2_{21},
\pi + \delta, a). 
\label{DeltaPflip2}
\end{equation}
Thus, $\Delta P_{flip} = \Delta P_{CP-CP}$; they are identical.

\section{Leading-order corrections to the CP-CP and the T-CP relations}

We now compute the leading-order corrections to the CP-CP and 
the T-CP relations. During the course of the computation, we will 
give an explicit proof of these relations. 
We start from the Kimura-Takamura-Yokomakura (KTY) formula 
\cite{KTY} of the oscillation probability, Eq.~(\ref{kty}).
We note that the coefficients $A$, $B$, and $C$ are functions of 
$\Delta m^2_{21}$, $\Delta m^2_{31}$, and the matter coefficient 
$a$, but we here suppress dependences on the latter two 
quantities. We also note that $A$ and $B$ start from first 
order in $\Delta m^2_{21}$, so that we can write 
$A(x) = x \alpha(x)$ and $B(x) = x \beta(x)$. 
Using the fact that 
$\sin(\pi+\delta)=-\sin(\delta)$ and 
$\cos(\pi+\delta)=-\cos(\delta)$, we obtain 
\begin{eqnarray}
\Delta P_{CP-CP} &\equiv& 
P(\Delta m^2_{21}, \delta) - P(-\Delta m^2_{21}, \pi+\delta),  
\nonumber \\
&=&
[A(\Delta m^2_{21})+A(-\Delta m^2_{21})]\cos \delta +
[B(\Delta m^2_{21})+B(-\Delta m^2_{21})]\sin \delta 
\nonumber \\
&+&
[C(\Delta m^2_{21})-C(-\Delta m^2_{21})],
\nonumber \\
&=&
[\alpha(\Delta m^2_{21}) - \alpha(-\Delta m^2_{21})] 
\Delta m^2_{21} \cos \delta +
[\beta(\Delta m^2_{21}) - \beta(-\Delta m^2_{21})] 
\Delta m^2_{21} \sin \delta 
\nonumber \\
&+&
[ C(\Delta m^2_{21}) - C(-\Delta m^2_{21})]. 
\label{DeltaCP3}
\end{eqnarray}
Therefore, we have shown that the RHS of (\ref{DeltaCP3}) is of order 
$\epsilon \equiv \frac{\Delta m^2_{21}}{\Delta m^2_{31}}$ ($C$ term), 
or $\epsilon^2$ ($A$ and $B$ terms). This is an explicit proof of 
the CP-CP relation, and hence also the T-CP  relation.

We are now left with the computation of the first-order terms 
of $\alpha$, $\beta$, and $C$. The exact form of 
these coefficients are calculated in \cite{KTY}.\footnote{
Note, however, that there is an error in the sign of the term denoted as $A^{(1)}_{k}$ 
in Eq.~(44) of the first reference in \cite{KTY}.
}
Therefore, it is straightforward to compute the RHS of (\ref{DeltaCP3}). 
It reads 
\begin{eqnarray}
\Delta P_{CP-CP} &=&
+ 16 
\frac{a \Delta m^2_{21} \Delta m^2_{31}}{(\Delta m^2_{31}-a)^3}
s_{13}^2 c_{13}^2 s_{12}^2 s_{23}^2 
\sin^2 \biggl[\frac{(\Delta m^2_{31} - a) L}{4E} \biggl] \nonumber \\
&-& 8 
\biggl( \frac{\Delta m^2_{21} L}{4E} \biggl) 
\biggl(\frac{\Delta m^2_{31}}{\Delta m^2_{31}-a}\biggl)^2
s_{13}^2 c_{13}^2 s_{12}^2 s_{23}^2  
\sin \biggl[\frac{(\Delta m^2_{31} - a) L}{2E} \biggl] 
\nonumber \\
&-& 16 J_{r} 
\biggl(\frac{\Delta m^2_{21}}{\Delta m^2_{31}-a}\biggl)^2
\Biggl[
\biggl(\frac{\Delta m^2_{31}}{a}\biggl)^2
(s_{12}^2-c_{12}^2) + 
\biggl(\frac{\Delta m^2_{31}}{a}\biggl)
(c_{12}^2-s_{12}^2) 
- s_{12}^2 
\Biggl]
\nonumber \\
&& \hspace*{35mm}\times
\sin\biggl(\frac{aL}{4E}\biggr)
\sin \biggl(\frac{(\Delta m^2_{31} - a) L}{4E} \biggr) 
\cos\biggl(\delta - \frac{\Delta m^2_{31} L}{4E} \biggr) 
\nonumber \\
&+& 8 J_{r}
\biggl( \frac{\Delta m^2_{21} L}{4E} \biggl) 
\biggl(\frac{\Delta m^2_{21}}{a}\biggl)
\biggl(\frac{\Delta m^2_{31}}{\Delta m^2_{31}-a}\biggl)
\Biggl[(s_{12}^2-c_{12}^2)
\sin \biggl(\delta - \frac{aL}{2E}\biggl) 
\nonumber \\
&&\hspace*{26mm} 
+ c_{12}^2
\sin\biggl(\delta - \frac{\Delta m^2_{31} L}{2E} \biggr) 
- s_{12}^2
\sin\biggl(\delta - \frac{(\Delta m^2_{31} - a) L}{2E} \biggl) \Biggl] 
\label{DeltaCPfinal}
\end{eqnarray}
where $J_{r} \equiv c_{12} s_{12} c_{23} s_{23} c_{13}^2 s_{13}$. 
%
$\Delta P_{T-CP}$ can be obtained by replacing 
$\delta$ by $2\pi - \delta$ 
in $\Delta P_{CP-CP}$, as dictated in (\ref{DeltaCPT}). 
We have kept in the expression of the oscillation probability 
the terms up to order 
$O(\epsilon s^2_{13})$ and $O(s^4_{13})$ in $C$, and to  
$O(\epsilon^2 s_{13})$ and $O(\epsilon s^3_{13})$ in 
$A$ and $B$. But the contributions from terms higher order 
in $s_{13}$ cancel in (\ref{DeltaCPfinal}).

The feature that the coefficients $A$ and $B$ start with 
first-order terms of $\Delta m^2_{21}$ played an important r{\^o}le 
in proving the CP-CP relation to leading order. 
It comes from the fact that they vanish in the two flavor limit 
$\Delta m^2_{21} \rightarrow 0$ 
and that the probabilities allow Taylor expansion in terms of 
the variable. The former statement is proved in Appendix 
on very general ground without assuming adiabaticity or 
constant matter density.

\section{Concluding remarks}

In this contribution to the Focus Issue on 'Neutrino Physics' 
we have presented a new unified view of the leptonic CP and T 
violation in neutrino oscillation. 
Based on the CPT relation in matter and other relations obeyed 
by the oscillation probabilities which are derived in \cite{MNP1} 
we were able to complete our understanding of the structure 
of unified description of CP and T violation in terms of 
bi-probability plot. 
Namely, the diamond shaped structure of simultaneous 
$P$- $CP[P]$ and $P$- $T[P]$ bi-probability plot is now 
understood as a consequence of a unique relation, the 
CP-CP (or the equivalently, the T-CP) relation. 
Based on this observation and relying on the KTY formula 
we have computed leading order corrections to the CP-CP relation.

We have also briefly touched upon the basic features of 
the T violating measure which are in contrast with those 
of CP violation. They include vanishing T violating measure 
at vanishing CP phase $\delta$, and the enhancement 
of T violating asymmetry by the matter effect up to a factor of 1.5.
Though measurement of T violation should give us a cleaner way 
of detecting genuine CP violating effects, it is not easy to 
carry out experimentally. We must wait for the construction of 
an intense electron (anti-) neutrino beam either by
beta beam \cite{beta} or in neutrino factories \cite{nufact}.

\begin{acknowledgments}

We thank Hiroshi Nunokawa and Stephen Parke for kindly redrawing 
the figure and the useful comments. 
MB wishes to thank the Sweden Japan Foundation for financing 
his visit to Tokyo Metropolitan University.
This work was supported by the Grant-in-Aid for Scientific Research
in Priority Areas No. 12047222, Japan Ministry
of Education, Culture, Sports, Science, and Technology, and 
the Grant-in-Aid for Scientific Research, No. 16340078, 
Japan Society for the Promotion of Science.

\end{acknowledgments}

\appendix
\section{No $\delta$-dependence in the two-flavor limit}

Though it should be the case on physics ground, it is not 
entirely trivial to show that $\delta$-dependence disappears 
from the oscillation probabilities 
in the two-flavor limit $\Delta m^2_{21} \rightarrow 0$. 
We carry it out explicitly in this Appendix. 
It is a slight modification of the method \cite{solarCP}
that allows us to show that $\delta$-dependence disappears 
in the survival probability $P(\nu_e \rightarrow \nu_{e})$.

We write down the evolution equation of three
flavor neutrinos in matter which is valid to leading order in
electroweak interaction:
\begin{equation}
i\frac{d}{dx}
\left[
\begin{array}{c}
\nu_e \\ \nu_\mu \\ \nu_\tau
\end{array}
\right]
=
\frac{1}{2E}
\left\{U \left[
\begin{array}{ccc}
0 & 0 & 0 \\
0 & \Delta m^2_{21} & 0 \\
0 & 0 & \Delta m^2_{31}
\end{array}
\right] U^{+}
+
\left[
\begin{array}{ccc}
a(x) & 0 & 0 \\
0 & 0 & 0 \\
0 & 0 & 0
\end{array}
\right]\right\}
\left[
\begin{array}{c}
\nu_e \\ \nu_\mu \\ \nu_\tau
\end{array}
\right].
\label{evolution1}
\end{equation}
In this Appendix we take a slightly different parametrization 
of the mixing matrix
\begin{equation}
U = e^{i\lambda_7 \theta_{23}} \Gamma_{\delta}e^{i \lambda_5\theta_{13}}
e^{i\lambda_2\theta_{12}}
\label{CKM}
\end{equation}
where $\lambda_i$ are $SU(3)$ Gell-Mann's matrix and $\Gamma$
contains the CP violating phase
\begin{equation}
\Gamma_{\delta} =
\left[
\begin{array}{ccc}
1 & 0 & 0 \\
0 & 1 & 0 \\
0 & 0 & e^{i \delta}
\end{array}
\right].
\end{equation}
We then rewrite the evolution equation (\ref{evolution1}) in terms of
the new basis defined by
\begin{eqnarray}
\tilde{\nu_{\alpha}} &=& \left[
\Gamma^{-1} e^{-i\lambda_7\theta_{23}}
\right]_{\alpha\beta} \nu_{\beta}
\nonumber \\
&\equiv& (T^t)_{\alpha\beta}\nu_{\beta}.
\label{defT}
\end{eqnarray}
In vanishing $\Delta m^2_{12}$ limit it reads
\begin{equation}
i\frac{d}{dx}
\left[
\begin{array}{c}
\tilde{\nu_e} \\ \tilde{\nu_\mu} \\ \tilde{\nu_\tau}
\end{array}
\right]
=
\frac{1}{2E}
\left\{e^{i\lambda_2 \theta_{13}} \left[
\begin{array}{ccc}
0 & 0 & 0 \\
0 & 0 & 0 \\
0 & 0 & \Delta m^2_{31}
\end{array}
\right] e^{-i \lambda_2\theta_{13}}
+
a(x)
\left[
\begin{array}{ccc}
1 & 0 & 0 \\
0 & 0 & 0 \\
0 & 0 & 0
\end{array}
\right]\right\}
\left[
\begin{array}{c}
\tilde{\nu_e} \\ \tilde{\nu_\mu} \\ \tilde{\nu_\tau}
\end{array}
\right]
\label{evolution2}
\end{equation}

Now we observe that the 
CP phase $\delta$ disappears from the equation. It is due to
the specific way that the matter effect comes in; $a(x)$ only
appears in (1.1) element in the Hamiltonian matrix and therefore
the matter matrix diag($a$, 0, 0) is invariant under rotation in
2-3 space by $e^{i\lambda_{7}\theta_{23}}$. Then the rotation by
the phase matrix $\Gamma$ does nothing.
Notice that $\tilde{\nu}_{\mu}$ does not have time evolution 
due to (\ref{evolution2}).

It is clear from (\ref{evolution2}) that any transition amplitudes
computed with $\tilde{\nu}_{\alpha}$ basis is independent of the CP
violating phase.
Of course, it does not immediately imply that the CP violating phase
$\delta$ disappears in the physical transition amplitude $\langle
\nu_{\beta} \mid \nu_{\alpha} \rangle$. The latter is related with
the transition amplitude defined with $\tilde{\nu}_{\alpha}$ basis as
\begin{equation}
\langle \nu_{\beta} \mid \nu_{\alpha} \rangle
=
T_{\alpha\gamma} T^*_{\beta\delta}
\langle \tilde{\nu_{\delta}} \mid \tilde{\nu_{\gamma}} \rangle
\end{equation}
where $T$ is defined in (\ref{defT}) and its explicit form in our
parametrization (\ref{CKM}) of the mixing matrix reads
\begin{equation}
T =
\left[
\begin{array}{ccc}
1 & 0 & 0 \\
0 & c_{23} & s_{23}e^{i\delta} \\
0 & -s_{23} & c_{23}e^{i\delta}
\end{array}
\right]
\end{equation}

One can show that
the amplitude of $\nu_e \rightarrow \nu_{\mu}$ has a
pure phase factor $e^{i\delta}$ and hence
$P(\nu_e \rightarrow \nu_{\mu})$ is independent of phase $\delta$;
\begin{equation}
\langle \nu_{\mu}(x) \mid \nu_e(0) \rangle
=
c_{23}\langle
\tilde{\nu}_{\mu} (x) \mid \tilde{\nu}_{e}(0) \rangle
+ s_{23}e^{i\delta}\langle
\tilde{\nu}_{\tau} (x) \mid \tilde{\nu}_{e}(0) \rangle
\end{equation}
The first term, however, vanishes because
$\langle\tilde{\nu}_{\mu} (x) \mid \tilde{\nu}_{e}(0)\rangle=
\langle\tilde{\nu}_{\mu} (0) \mid \tilde{\nu}_{e}(0)\rangle=0$.
(No evolution in $\tilde{\nu}_{\mu}$.)
Notice that the same statement does apply to the
$P(\nu_e \rightarrow \nu_{\tau})$ and
$P(\nu_{\mu} \rightarrow \nu_{\tau})$ as well.
One can show that the same conclusion holds for different 
choice of the phase matrix from that in (\ref{CKM}).

Since absence or presence of T violation should not depend on 
the parametrization used, this completes the proof that the 
$\delta$ dependence disappears from all the oscillation probabilities in the limit
$\Delta m^2_{12} \rightarrow 0$.


\end{document}